\def\gcmm3{{\,{\rm g\,cm^{-3}}}}

\def\fun#1#2{\lower3.6pt\vbox{\baselineskip0pt\lineskip.9pt
  \ialign{$\mathsurround=0pt#1\hfil##\hfil$\crcr#2\crcr\sim\crcr}}}
\documentstyle[preprint,prd,aps,epsfig,floats,graphics]{revtex}
\draft

\begin{document}

\tighten

\title{
\vskip1.55cm
A theory of thin shells with orbiting constituents}

\author{Victor Berezin}
\address{Institute for Nuclear Research of the Russian Academy of Sciences,\\
60th October Anniversary Prospect, 7a, 117312, Moscow, Russia\\
and\\
TH Division, CERN, Geneva, Switzerland\\
e-mail: berezin@ms2.inr.ac.ru}

\author{Maxim Okhrimenko}
\address{Institute for Nuclear Research of the Russian Academy of Sciences,\\
60th October Anniversary Prospect, 7a, 117312, Moscow, Russia\\
e-mail: okhrim@pcbai10.inr.ruhep.ru}

\maketitle

\begin{abstract}
{\it
The self-gravitating, spherically symmetric thin shells built of orbiting 
particles are studied. Two new features are found. One is the minimal 
possible value for an angular momentum of particles, above which elliptic 
orbits become possible. The second is the coexistence of both the wormhole 
solutions and the elliptic or hyperbolic orbits for the same values of the 
parameters (but different initial conditions). Possible applications of these 
results to astrophysics and quantum black holes are briefly discussed.
}

\end{abstract}

\section{Introduction}
Thin shells play an important role in General Relativity. Because of 
the non-linearity of Einstein's equations we have to consider some simplified 
models that, describing essential features of the problem under 
consideration, can be solved analytically at least in quadrature. The thin 
shell formalism provides us with such a source of the gravitational field, 
the first integrals being incorporated in the very structure of the 
Einstein equations for the shells.

This formalism was first developed in \cite{Isr} and applied to the 
gravitational collapse problem \cite{Frolov}. It was then almost forgotten 
for many years. Only in 1983 was it revived and used for studying dynamics 
of vacuum phase transitions in the Early Universe \cite{3,4,5,6,7,8,9,10}. 
The possibility to obtain self-consistent solutions to the gravity 
equations together with the matter source equations has led to very 
interesting results (restrictions on the parameters of a vacuum decay 
process, enhancing of a vacuum decay probability around black holes and 
others) \cite{3,6,9,10,11,bkt8}, which were impossible to obtain by studying  
the dynamics of test particles (or fields) in a given background space-time.

Comparing the self-consistent treatment of gravitating matter sources 
and the motion of the same source in a given gravitational field we 
should stress that the main advantage of the former is that in this case 
we obtain an important information about a global space-time structure 
while in the latter case the matter fields ``feel'' space-time only locally.

The very possibility to feel the space-time globally is extremely important 
in quantum theory. It was recently demonstrated \cite{bbn} in quantum black 
hole models by discovering a new quantum number. Thus, thin shell models serve 
as a powerful tool in the investigation of a variety of problems where effects 
of strong gravitational field may prove essential.

In the present paper we studying a new class of spherically symmetric, 
self-gravitating thin shells. Namely, these shells are built of particles 
that follow elliptic or hyperbolic orbits, their pericentre being smeared 
uniformly upon the sphere with zero average angular momentum. The paper is
organized as follows. In the Section II we briefly introduce the thin
shell formalism with the emphasize on the spherically symmetric case. 
In the Section III the surface energy-momentum tensor is derived 
for the spherically symmetric shells with orbiting constituents. 
The Section IV is devoted to the dynamics of such shells. In the Section V we 
consider a special, very important case of light-like particles. 
And, finally, in the Section VI we discussed the obtained results and their 
application to astrophysics and quantum black holes.

Throughout the paper we use the units $\hbar =c=1$ ($\hbar$ is the
Planckian constant and $c$ is the speed of light). In these units the
Planckian mass $m_{Pl}\simeq 10^{-5}\,$gr, length $l_{Pl}\simeq 10^{-33}$cm
and time $t_{Pl}\simeq 10^{-43}$s are expressed in terms of the Newtonian
constant $G$ as $G=m_{Pl}^{-2}=l_{Pl}^{2}=t_{Pl}^{2}$.

\section{General preliminaries}

1. Spherically symmetric space-times

A general spherically symmetric space-time is a direct product of a 
two-dimensional sphere $S^{2}$ and a two-dimensional pseudo-Euclidean 
space-time $M^{2}$. Therefore, a general line element can be written in the 
form 
\begin{equation}
ds=g_{\mu \nu }dy^{\mu }dy^{\nu }=\gamma _{ik}dx^{i}dx^{k}-R^{2}(x)d\sigma
^{2}\,.
\label{1}
\end{equation}
 
Here $g_{\mu \nu }$ is a metric tensor of a four-dimensional space-time 
$M^{4}$, $y^{\mu }$ are the coordinates on $M^{4}$, $\gamma _{ik}$ is a metric 
tensor of $M^{2}$ in a coordinate system $x^{i}$, $R(x)$ is a radius of a 
two-dimensional sphere and \ $d\sigma ^{2}$ is the line element of $S^{2}$,
which can be parametrized with two angles, $\theta $ and $\varphi $: 
\begin{equation}
d\sigma ^{2}=d\theta ^{2}+\sin ^{2}\theta d\varphi ^{2}  .
\label{2}
\end{equation}

The radius $R(x)$ is defined in such a way that the area of the sphere 
equals $4\pi R^{2}(x)$. Since Einstein's equations of General Relativity are 
invariant under general coordinate transformation we are able to subject 
altogether three components of two-dimensional metric $\gamma _{ik}$ to two 
conditions. Thus, a general spherically symmetric space-time can be locally 
described by only two functions: one is just the radius $R(x)$ and another 
one comes from $M^{2}$. It can be shown (for details see \cite{9,lect}) that 
for this second function we can choose an invariant 
\begin{equation}
\Delta =\gamma ^{ik}R_{,i}R_{,k}  ,
\label{3}
\end{equation}where $\gamma ^{ik}$ is the inverse to the metric tensor 
$\gamma 
_{ik}$, $%
R_{,i}$ is a partial derivative of the radius with respect to a coordinate 
$%
x^{i}$. The invariant function $\Delta $ is nothing but the square of 
the vector normal to the surfaces of constant radius $R(x)=$const.

 This normal vector can be space-like ($\Delta <0$), time-like ($%
\Delta >0$) or null ($\Delta =0$). It is simple to analyse these 
possibilities using orthogonal coordinates. Choosing some time coordinate 
$t$ 
and some radial coordinate $q$, we have 

\begin{equation}
\Delta =A\dot{R}^{2}-B {R^{\prime }}^{2},\,A(t,q),B(t,q)>0  ;
\label{4}
\end{equation}
here dot and prime stand for derivatives along time ($t$) and radial ($q$) 
directions respectively. In the case of negative $\Delta $ the surfaces of 
constant radius are time-like. We have already used one of the two possible 
coordinate transformations (gauge freedom) to make coordinates orthogonal. 
Now we can exploit the remaining freedom to put $q=R$. The region where the 
radius can be chosen as a radial coordinate is called an $R$-region. 
Similarly, for positive $\Delta $ and space-like surfaces of constant radius 
we can put $t=R$. The region where the radius can be chosen as a time 
coordinate is called a $T$-region. The notions of $R$- and $T$-regions were
introduced in \cite{Novik}.

 Further, in the $R$-region ($\Delta <0$) we can never have $R^{\prime
}=0$. So the sign of a partial derivative of the radius is an invariant. So, 
we have either $R^{\prime }>0$, which is called the $R_{+}$-region, 
or $R^{\prime }<0$, called the $R_{-}$-region. If we assume 
that a radial coordinate $q$ ranges from $-\infty $ at the left to $+\infty $ 
at the right and call inner region the region to the left of the given 
surface $R=$const. and outer region the one to the right, then, in 
a $R_{+}$-region, the radii increase outside this given sphere, while they 
derease in the $R_{-}$-region. In the flat Minkowsky space-time $\Delta =-1$, 
so we have globally 
an $R_{+}$-region and, can thus choose $q=R$ everywhere.

 Analogously, in a $T$-region ($\Delta >0$) we can never have $%
\dot{R}=0$. Thus, there may be regions of inevitable expansions with $%
\dot{R}\,>0$, and regions of inevitable contraction with $\dot{%
R}\,<0$. The former are called $T_{+}$-regions, while the latter are $T_{-}$. 

 The null surfaces of constant radius ($\Delta =0$) are called the 
apparent horizons and serve as boundaries between $R$- and $T$-regions.

 Thus, spherically symmetric space-time may have quite rich a structure. 
In general, it is some set of $R_{\pm }$ and $T_{\pm }$-regions 
separated by the apparent horizons. The most famous (and most important 
to us) example is the geodesically complete Schwarzschild space-time. 
The space-time is geodesically complete (or maximal analytically extended) 
if all the geodesics start and end either at infinities or at singularities. 
The Carter--Penrose conformal diagram for the Schwarzschild manifold is shown 
in Fig. 1 (on the conformal diagram all the infinities are brought to final 
distances).
\begin{figure}
\begin{center}
\epsfig{file=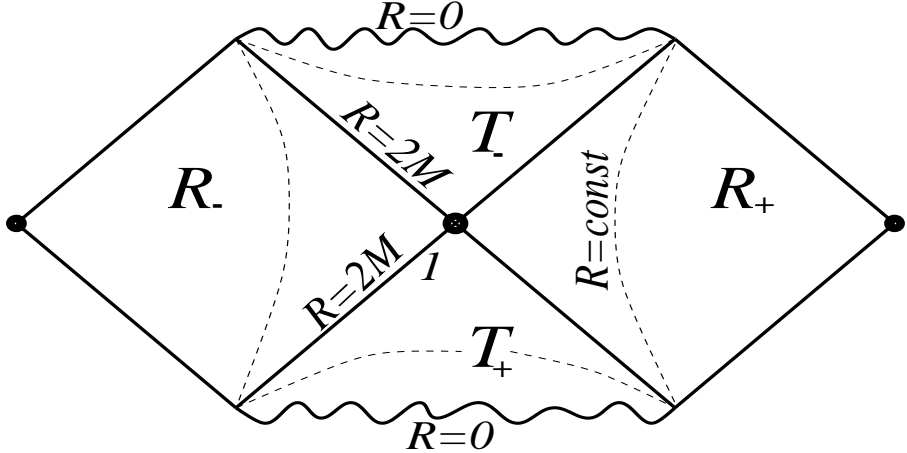,width=12cm,height=6cm}
\end{center}
\caption{Carter--Penrose diagram for the maximally extended Schwarzschild 
space-time.
\label{carter} }
\end{figure}

We see that in this simple case all possible regions are 
present. We have two isometric $R$-regions ($R_{-}$ to the left and $R_{+}$ 
to the right), each of them having an infinity, and two isometric $T$%
-regions ($T_{+}$-region of inevitable expansion in the past and $T_{-}$%
-region of inevitable contraction in the future, $T_{+}$-region starting 
and $T_{-}$-region ending with real singularities at $R=0$, which are 
space-like surfaces).

2. Thin shell formalism

 In General Relativity the inclusion of matter sources into considerations 
is not an easy task because of nonlinearity of Einstein's equation. Besides, 
the investigation of a self-consistent solution of a combined system of 
gravity and matter evolution is extremely important since it can differ 
substantially (and crucially) from the evolution of the same matter in a given 
background field. Thus there is a necessity in the construction of a very 
simple and tractable model for matter sources. One of these model are thin 
shells. The thin shells are viewed as a vanishing limit of thick shells of 
ordinary matter. In some cases, e.g. in cosmological phase transitions, such 
a limit can be justified \cite{3}. 

The general mathematical theory of thin shells in General Relativity was 
invented in \cite{Isr}. It was formulated in a nice geometrical way as the set 
of equations for an extrinsic curvature tensor, which describes the embedding 
of a three-dimensional hyper-surface (world-sheet of a thin shell) into 
four-dimensional space-time (of course, a number of dimensions can be made 
arbitrary $(d-1)$\ and $d$). Since a non-zero amount of energy is concentrated 
in a vanishing volume, such a thin shell is singular and, while all the 
metric coefficients are continuous across a singular hyper-surface, some of 
their derivatives undergo a jump. The equations for this jump are the 
following \cite{9} 
\begin{equation} 
\lbrack K_{i}^{j}]-\delta _{i}^{j}[K_{l}^{l}]=8\pi GS_{i}^{j}  , 
\label{5}
\end{equation}
where $K_{i}^{j}$ is the extrinsic curvature tensor describing the way of 
embedding of a time-like three-dimensional hyper-surface into a 
four-dimensional space-time (for a space-like hyper-surface one should change 
the sign of the right (or left) hand side). The square brackets denote a 
jump across a hyper-surface, $[A]=A_{out}-A_{in}$, $S_{i}^{j}$ is a surface 
energy momentum tensor of a shell. From the Einstein's equations just outside 
the shell there follow two more expressions  
\begin{equation}
S_{i\mid \,j}^{j}+[T_{i}^{n}]=0
\label{6}
\end{equation}
\begin{equation}
\{K_{j}^{i}\}S_{i}^{j}+[T_{n}^{n}]=0  .
\label{7}
\end{equation}
Here $\{K_{j}^{i}\}=\frac{1}{2}(K_{j}^{i}$(out)$-K_{j}^{i}$(in)), and the 
index $n$ denotes the outward direction normal to the shell. 

In the spherically symmetric case everything becomes much easier. Of the 
total set of Eq.(\ref{5}) we are left with only two: 
\begin{equation}
\lbrack K_{2}^{2}]=4\pi GS_{0}^{0}
\label{8}
\end{equation}
\begin{equation}
\lbrack K_{0}^{0}]+[K_{2}^{2}]=8\pi GS_{2}^{2}\,
\label{9}
\end{equation}
and from Eqs.(\ref{6}) and (\ref{7}) we again have only two equations 
\begin{equation}
\{K_{0}^{0}\}S_{0}^{0}+2\{K_{2}^{2}\}S_{2}^{2}+[T_{n}^{n}]=0
\label{10}
\end{equation}
\begin{equation}
\dot {S_{0}^{0}}+\frac{2\dot{\rho }}{\rho }%
(S_{0}^{0}-S_{2}^{2})+[T_{0}^{n}]=0  .
\label{11}
\end{equation}
Here a dot means a derivative with respect to the proper time of the shell. 
Moreover, it appears \cite{9}, that Eq.(\ref{10}) is an algebraic 
consequence of 
other equations, and Eq.(\ref{11}) can be considered as an integrability 
condition to Eqs.(\ref{8}) and (\ref{9}). Therefore, we can use any two of 
three 
equations, Eqs.(\ref{8}), (\ref{9}) and (\ref{11}). Another advantage of 
spherical symmetry is 
that we are able to calculate a $K_{2}^{2}$-component of the extrinsic 
curvature tensor for a general spherical space-time \cite{9}. The result is  
\begin{equation}
K_{2}^{2}=-\frac{\sigma }{\rho }\sqrt{\dot{\rho }^{2}-\Delta\,}  ,
\label{12}
\end{equation}
where $\rho $ is the radius of the shell as a function of the proper time, 
$\dot{\rho }$ is its derivative, and $\Delta $ is an invariant function 
for a given spherical space-time, introduced earlier. What is $\sigma $? It 
is a sign function: $\sigma =+1$ if the radii increase just outside the 
shell (in the outward normal direction) and $\sigma =+1$ if they decrease. 
We know already that $\sigma =+1$ in any $R_{+}$-region, and $\sigma =-1$ in 
any $R_{-}$-region. The sign of $\sigma $ can change only in $T_{-}$-regions. 

At the end of this section we would like to make a rather important note. It 
concerns a qualitative difference between two approaches to the 
matter-gravitation dynamics. In the first approach one considers matter 
motion in a 
given background (say, Schwarzschild) metric. In the case of dust particles 
it means simply a geodesic motion. In the second case the motion is 
self-consistent, i.e. the full back-reaction of the matter fields on the 
space-time metric is taken into account. This is best 
illustrated by some simple model. Let us consider a spherically symmetric 
thin shell consisting of radially moving dust particles with total bare mass 
$M$; then $S_{0}^{0}=\frac{M}{4\pi \rho ^{2}}$. Let them move in the 
gravitational field of a black hole of mass $m$. Owing to the back-reaction, 
the total Schwarzschild mass outside the shell will be, say, $m+\Delta m$, $%
\Delta m>0$. Our main equation (\ref{8}) is now 
\begin{equation}
\sqrt{\dot {\rho }^{2}+1-\frac{2Gm}{\rho }}-\sigma \sqrt{\dot {%
\rho }^{2}+1-\frac{2G(m+\Delta m)}{\rho }=}\frac{GM}{\rho } ,
\label{13}
\end{equation}
where $\sigma =\sigma _{out}$ ($\sigma _{in}=1$ because we put the shell 
outside the black hole). For the sake of simplicity we are interested only 
in a bound motion. Let $\rho _{0}$ be a turning point, then 
\begin{equation}
\sqrt{1-\frac{2Gm}{\rho _{0}}}-\sigma \sqrt{1-\frac{2G(m+\Delta m)}{\rho _{0}%
}=}\frac{GM}{\rho _{0}} .
\label{14}
\end{equation}
If $\sigma =+1$, the turning point is in the $R_{+}$-region of the outer 
metric and we call this a black hole case. For $\sigma =-1$, 
$\rho _{0}$ lies in the  
$R_{-}$-region and we have a wormhole case. Let us introduce the following 
convenient notation $\varepsilon =\frac{m}{M}$ and $\Delta \varepsilon =%
\frac{\Delta m}{M}$. It can be easily shown that 
\begin{eqnarray}
\sigma =+1\qquad    {\rm{if}}\qquad    \Delta \varepsilon >\frac{1}{2}(\sqrt{%
\varepsilon ^{2}+1}-\varepsilon )
\\
\sigma =-1\qquad    {\rm{if}}\qquad    \Delta \varepsilon <\frac{1}{2}(\sqrt{%
\varepsilon ^{2}+1}-\varepsilon )  .\nonumber
\end{eqnarray}
For $\varepsilon \gg 1$ we have 
\begin{eqnarray}
\sigma =+1\qquad {\rm{if}}\qquad \Delta \varepsilon >\frac{1}{4\varepsilon }
\\
\sigma =-1\qquad {\rm{if}}\qquad \Delta \varepsilon <\frac{1}{4\varepsilon } .
\nonumber
\end{eqnarray}
In the Schwarzschild background limit $\Delta \varepsilon \gg \varepsilon $ 
we are left with only the $R_{+}$-region. Of course the shell can move in 
the $R_{-}$-region as well, but in this limit the observer cannot distinguish 
between 
``in'' and ``out'' (left and right on the Carter--Penrose diagram), so the 
shell can fill only local geometry. But with a self-consistent description we 
are able to probe a global geometry as well, and this may have far-going 
consequences for a quantisation procedure \cite{bbn}. 

\section{Thin shells with orbiting constituents}

1. Construction.

Our aim is to study the dynamics of a self-gravitating thin shells with 
orbiting constituents. What does the expression orbiting constituents mean? 
Let us consider a 
point mass moving according to law of  Newtonian gravity (Kepler's problem 
in Celestial mechanics) in a field of gravitating centre. Its orbit is an 
ellipse with such a centre at one of the focuses. If we build an ensemble of 
such particles with the same angular momentum-to-mass ratio, they will have 
the 
same value of both pericentre and apocentre. Let us imagine that, initially, 
all 
these particles are smeared uniformly on a surface of a sphere whose 
radius is that of pericentre and let them start simultaneously (with equal 
absolute value of the velocity but in different directions, i.e. in different 
planes). Then such an ensemble will form a spherically symmetric thin shell 
oscillating between a pericentre and an apocentre. This is exactly what we 
call a thin shell with orbiting constituents. Such a construction can be 
applied equally to both non-relativistic and relativistic Coulomb problems. Of 
course,in relativistic case, orbits are no longer closed ellipses, but 
it is all the same qualitatively. 

We confine ourselves to the spherically symmetric case, not only because it 
is simple to treat. The equations for the gravitational field are non-linear, 
moreover, the back-reaction of the matter fields plays an important role in 
the case of strong gravitational fields, e.g. in the black hole physics we 
are mostly interested in, so that the addition of one source to another does 
not 
lead to a simple sum of the resulting gravitational fields. Besides, our 
distant goal is the construction of a quantum theory of such shells and we 
want to 
avoid the complications caused by the gravitational waves and, thus,
encountering all 
the difficulties and obstacles of the full quantum gravity. These 
unpleasant future are absent in the spherically symmetric models. 

The shells with orbiting constituents can also be useful in astrophysics, 
e.g. 
in studying dynamics of globular clusters. The very idea to construct 
spherically symmetric shells out of orbiting stars belongs to 
Bisnovatyi-Kogan \cite{Kogan}. But the full application of the idea to the 
analytical and numerical investigation of globular clusters requires 
consideration of intersecting shells, and this problem is far from being 
solve. 

2. Surface energy--momentum tensor. 

We already saw that,for a full description of the thin shell in the case of 
spherical symmetry, we need only one equation, Eq.(\ref{8}) (which is nothing 
but the energy conservation constraint). In order to solve it we should 
know the surface energy density $S_{0}^{0}$ of our shell. This can be 
deduced from the integrability condition, Eq.(\ref{11}), provided we know an 
``equation of state'' $S_{2}^{2}=S_{2}^{2}(S_{0}^{0})$ or that we 
calculate it independently. Before going to calculations, some general words 
are in order. 

The energy and the energy density depend, of course, on the choice of the 
time coordinate. In our case we need the surface energy density measured by 
an observer sitting on the shell (and not looking around). For this observer, 
the choice of the time coordinate is very poor, namely, $t=t(\tau )$ 
where $t $ is the proper time. But under such a reparametrization 
$S_{0}^{0}$ is 
invariant (unlike $S^{0\,0}$\ or $S_{0\,0}$). This invariance can also be
understood in the following way. The angular components of the surface 
stress--energy tensor $S_{2}^{2}=S_{3}^{3}$ are invariant under 
transformations in a factorized two-dimensional space-time $M_{2}$. The trace  
Tr$\,S_{j}^{\,i}$\ is also an invariant, and so is the mixed temporal 
component $S_{0}^{0}=$Tr$S_{j}^{\,i}-S_{2}^{\,2}-S_{3}^{3}=$Tr$S_{j}^{\,i}-
2\,S_{2}^{\,2}$. Thus, $S_{0}^{0}$ can depend only on the invariant functions 
of an underlying space-time ($R$\ and $\Delta $). Moreover, $S_{0}^{0}$ 
should not 
depend on $\Delta $ because in a two-dimensional space-time $M_{2}$ a 
world-sheet of our thin shell becomes a world-line, and by appropriate 
coordinate transformation any space-time can be made flat along the line 
(irrespective of the fact that $S_{0}^{0}$ describes the internal structure 
of the shell and not its embedding properties). Thus, $S_{0}^{0}$ depends 
only on the radius $R$ of the shell and some appropriate integrals of 
motions.
 
In order to get a surface energy density $S_{0}^{0}$ we first calculate the 
sum of energies of the constituent particles in the frame of reference of an 
observer sitting on the shell. Then, taking into account that particles are 
uniformly distributed on the sphere of radius $\rho $ we just divide the 
energy by $4\pi \rho ^{2}$. To fulfil this programme let us consider a point 
particle of mass $\mu $. The action integral for such a particle is (see 
e.g. \cite{Landau})  
\begin{eqnarray} 
S &=&-\mu \int ds=(ds^{2}=g_{\alpha \beta }\,dy^{\alpha }dy^{\beta })=
\nonumber \\
&=&-\mu \int \sqrt{g_{\alpha \beta }\,du^{\alpha }du^{\beta }}ds=-\mu \int 
\sqrt{g_{\alpha \beta }\,dy^{\alpha }dy^{\beta }}= \\
&=&-\mu \int \sqrt{g_{\alpha \beta }\,\frac{dy^{\alpha }}{dt}\frac{dy^{\beta
}}{dt}}=\int Ldt  .\nonumber
\label{action}
\end{eqnarray}
Here $u^{\alpha }=\frac{dy^{\alpha }}{dt}$ is a four-velocity, $t$ is some 
time coordinate ($y^{0}=t$), and $L$ is a Lagrangian and, thus, the energy 
heavily depends on the choice of time. Since we are interested in the
spherically symmetric space-times seen by the observer sitting on the 
spherical shell the line element takes the form 
\begin{equation}
ds^{2}=d\tau ^{2}-\rho ^{2}(\tau )\,(d\theta ^{2}+\sin ^{2}\theta \,d\varphi
^{2}) ,
\label{ds}
\end{equation}
where $\rho $ is the radius as a function of the proper time $\tau $. Then, 
as usual, we find the momenta $p $ and the Hamiltonian (the dot means a 
derivative with respect to $\tau $): 
\begin{eqnarray}
L=-\mu \,\sqrt{1-\rho ^{2}\dot {\theta }^{2}-\rho ^{2}\dot {%
\varphi }^{2}} ,
\nonumber\\
p_{\theta }=\frac{\mu \,\rho ^{2}\sin ^{2}\theta \dot {\theta }}{\sqrt{%
1-\rho ^{2}\dot {\theta }^{2}-\rho ^{2}\sin ^{2}\theta \dot {%
\varphi }^{2}}}=\mu \,\rho ^{2}u^{\theta }=-\mu \,u_{\theta } ,
\nonumber\\
p_{\varphi }=\frac{\mu \,\rho ^{2}\sin ^{2}\theta \dot {\varphi }}{%
\sqrt{1-\rho ^{2}\dot {\theta }^{2}-\rho ^{2}\sin ^{2}\theta \dot
{\varphi }^{2}}}=\mu \,\rho ^{2}u^{\varphi }=-\mu \,u_{\varphi } ,
\end{eqnarray}
\begin{eqnarray}
H &=&p_{\theta }\dot {\theta }+p_{\varphi }\dot {\varphi }-L=\mu
\,u_{0}= \nonumber\\
&=&\mu \,\sqrt{1-u_{\theta }u^{\theta }-u_{\varphi }u^{\varphi }}=\sqrt{\mu
^{2}+\frac{1}{\rho ^{2}}p_{\theta }^{2}+\frac{1}{\rho ^{2}\sin ^{2}\theta }
p_{\varphi }^{2}} .\nonumber
\label{H}
\end{eqnarray}
Now, from the Hamiltonian equations 
\begin{eqnarray}
\dot {\theta }=\frac{\partial H}{\partial p_{\theta }}   ,\qquad   
\dot {\varphi }=\frac{\partial H}{\partial p_{\varphi }} ;
\\
\dot {p_{\theta }}=\frac{\partial H}{\partial \theta }   ,\qquad 
\dot {p_{\varphi }}=-\frac{\partial H}{\partial \varphi }\nonumber
\label{HE}
\end{eqnarray}
it follows that 
\begin{equation}
p_{\theta }^{2}+\frac{1}{\sin ^{2}\theta }p_{\varphi }^{2}={\rm{const}}=J^{\,2}
\label{f}
\end{equation}
where $J$ is the conserved angular momentum of each particle \cite{Landau}.
Hence, for the energy of our shell (remember that all particles have the
same ratio $a=\frac{J}{\mu }$) we obtain 
\begin{equation}
E=M\sqrt{1+\frac{a^{2}}{\rho ^{2}}} ,
\label{E}
\end{equation}
where $M$ is the bare mass of the shell (i.e. the sum of all the rest mass
of the constituent particles). Thus, for the surface energy density we have 
\begin{equation}
S_{0}^{0}=\frac{M}{4\pi \rho ^{2}}\sqrt{1+\frac{a^{2}}{\rho ^{2}}} .
\label{S}
\end{equation}
Having $S_{0}^{0}$ we can easily calculate $S_{2}^{2}=S_{3}^{3}$. Indeed, 
the structure of the surface energy--momentum tensor $S_{i}^{\,j}$ copies 
that of 
the energy-momentum--tensor $T_{\alpha }^{\,\beta }$ for the dust
matter 
\begin{eqnarray}
S_{i}^{\,j}=A\,u_{i}\,u^{j}
\nonumber\\
u_{i}\,u^{i}=1 ,
\label{SU}
\end{eqnarray}
except that now $\,u^{1}=0$ since we are sitting on the shell. Then, we have 
\begin{eqnarray}
u_{\theta }\,u^{\theta }+u_{\varphi }\,u^{\varphi }=-\frac{J^{2}}{\mu
^{2}\rho ^{2}}=-\frac{a^{2}}{\rho ^{2}}   ,
\nonumber\\
u_{0}\,u^{0}=1+\frac{a^{2}}{\rho ^{2}}   ,
\nonumber\\
S_{0}^{\,0}=\frac{M}{4\pi \rho ^{2}}\sqrt{1+\frac{a^{2}}{\rho ^{2}} }%
  ,
\\
A=\frac{M}{4\pi \rho ^{2}\sqrt{1+\frac{a^{2}}{\rho ^{2}}}}   ,
\nonumber\\
S_{2}^{\,2}=S_{3}^{\,3}=-\frac{M\,a^{2}}{8\pi \rho ^{4}\sqrt{1+\frac{a^{2}}{%
\rho ^{2}} }}   .\nonumber
\label{stress}
\end{eqnarray}
It can be readily checked that with these $S_{0}^{\,0}$ and$\ S_{2}^{\,2}$
the integrability condition inEq.(\ref{11}) is identically satisfied.

Now we have everything for analysing a dynamical evolution of a 
self-gravitating thin shell with orbiting constituents.

For further purposes we complete this section by writing the expression
for Hamiltonian and energy for our thin shell in the specific background
metric: 
\begin{equation}
ds^{2}=F\,dt^{2}-\frac{1}{F}dr^{2}-r^{2}(d\theta ^{2}+\sin ^{2}\theta
\,d\varphi ^{2})\,  .
\label{F}
\end{equation}
This form includes the flat Minkovski space-time in Lorentzian coordinates
for $F=-\Delta =1$, Schwarzschild space-time (i.e. a metric outside
a spherically symmetric, neutral, massive object of mass $m$) for 
$F=-\Delta =1-\frac{2Gm}{r}$, and Reissner--Nordstrom space-time (a metric 
outside a spherically symmetric, electrically charged, massive object 
of mass $m$ and
charge $e$) for $F=-\Delta =1-\frac{2Gm}{r}-\frac{G\,e^{2}}{r}$. The
Hamiltonian is 
\begin{equation}
H=\sqrt{F}\sqrt{\mu ^{2}+F\,p_{r}^{2}+\frac{J^{2}}{r^{2}}}   .
\label{HF}
\end{equation}
Note the appearance of the radial momentum $p_{r}$ absent, for the on-shell 
observer. The expression for the shell energy written in terms of the radial 
velocity ($\frac{dr}{dt}$)\ reads as follows 
\begin{equation}
E=\frac{F^{3/2}M}{\sqrt{F^{2}-\dot {r}^{2}}}\sqrt{1+\frac{a^{2}}{r^{2}}%
} .
\label{EF}
\end{equation}
Note that now it is not the energy measured by an on-shell-observer, but the 
total energy (including all interactions a kinetic energies) measured by a 
distant observer (sitting near infinity). We can rewrite it in terms of the 
proper time:

\begin{eqnarray}
Fdt^{2}-\frac{1}{F}dr^{2}=d\tau ^{2}
\nonumber\\
F-\frac{1}{F}\,\left(\frac{dr}{dt}\right)^{2}=
\left(\frac{d\tau }{dt}\right)^{2}
\nonumber\\
F\,\left(\frac{dt}{d\tau }\right)^{2}-\frac{1}{F}\dot {\rho }^{2}=1
\\
\frac{F^{3}}{F-(\frac{d\tau }{dt})^{2}}=\dot {\rho }^{2}+F
\nonumber\\
E=M\sqrt{\dot {\rho }^{2}+F}\sqrt{1+\frac{a^{2}}{\rho ^{2}}} .\nonumber
\label{Etan}
\end{eqnarray}
The last expression looks a little bit more elegant than Eq.(\ref{EF}). It 
is worth while to know that the same expressions for the energy follow from 
geodesic equations in the corresponding space-times, as it should be.

\section{Dynamics of thin shells with orbiting particles}

In order to understand how General Relativity changes the dynamics of shells 
with orbiting particles, we first consider the well-known relativistic
Coulomb problem, then we investigate the shells in the given Schwarzschild 
background and, at the last part of this Section, we show what is new that 
comes from accounting for a back-reaction of a shell dynamics on a space-time 
metric.

1. Relativistic Coulomb problem

To derive the main equation, which is just the energy conservation equation, 
we start with our general expression, Eq.(\ref{8}), for a charged thin shell 
with 
orbiting particles with $S_{0}^{0}$\ given by Eq.(\ref{S}). In this case 
the 
metric both inside and outside the shell is given, in general, by the 
Reissner--Nordstrom solution to the Einstein equation with 
\begin{equation}
F=-\Delta =1-\frac{2\,G\,m}{r}+\frac{G\,Q^{2}}{r^{2}} , 
\label{R-N}
\end{equation}
where $G$ is the Newton constant, $m$ is the total mass (energy) of the 
system, and $Q$ is its electric charge. We denote the mass and the charge 
inside the shell by $m$ and $Q$, respectively, and that of the outside 
region by the $(m+\Delta m)$\ and $(Q+q)$, respectively. Then, the main 
equation reads as follows (in this subsection we need consider only the 
case with $\sigma _{in}=\sigma _{out}=+1$): 
\begin{eqnarray}
\sqrt{\dot {\rho }^{2}+1-\frac{2\,G\,m}{\rho }+\frac{G\,Q^{2}}{\rho
^{2}}}-\sqrt{\dot {\rho }^{2}+1-\frac{2\,G\,(m+\Delta m)}{\rho }+
\frac{G(\,Q+q)^{2}}{\rho ^{2}}}= 
\nonumber\\
\frac{G\,M}{\rho }\sqrt{1+\frac{a^{2}}{\rho
^{2}}}  ,
\end{eqnarray}
where $M$ is the bare mass, $a$ is the specific angular momentum of each 
particle, $\rho (\tau )=r(t(\tau ))$ is the radius as a function of the 
proper time $\tau $ on the shell, and dots denote the time derivatives. 
Taking now the limit of vanishing gravity $(G\longrightarrow 0)$ we obtain 
\begin{equation}
\Delta m=\sqrt{\dot {\rho }^{2}+1}\sqrt{1+\frac{a^{2}}{\rho ^{2}}}+%
\frac{Q\,q+\frac{1}{2}\,q^{2}}{\rho }  , 
\label{coulomb}
\end{equation}
where $\Delta m$ is the total mass (energy) of the shell. For a hydrogen-like 
atom, $Q=Z\,e$, $q=-e$ (electron charge), and after changing notation for 
the total energy $(\Delta m\longrightarrow E)$ we have 
\begin{equation}
E=\sqrt{\dot {\rho }^{2}+1}\sqrt{1+\frac{a^{2}}{\rho ^{2}}}+\frac{%
Z\,e^{2}+\frac{1}{2}\,e^{2}}{\rho }  . 
\label{atom}
\end{equation}
This last equation differs from the conventional form of the relativistic 
Coulomb problem in two respects. First, the kinetic term is different 
because, as we already explained, we use the proper time of the shell rather 
than the Minkowskian coordinate time. Then, there is an additional repulsive 
term proportional to $e^{2}$, which describes the self-interaction of the 
charged spherical shell. Introducing a ``potential'' by 
$\Delta m(\dot {\rho }=0)=V(\rho )+1$,\ we get 
\begin{equation}
V(\rho )=M\sqrt{1+\frac{a^{2}}{\rho ^{2}}}-\frac{J_{cr}}{\rho }-M  , 
\label{potential}
\end{equation}
where by $J_{cr}=Z\,e^{2}-\frac{1}{2}e^{2}$ we denoted the critical value of 
the angular momentum. Remembering that $a=\frac{J}{\mu }$, we have in the 
case of a single orbiting particle $(M=\mu )$ two different curves for our 
potential $V(\rho )$ depending on whether $J<J_{cr}$\ or $J>J_{cr}$\ , as  
shown in Fig. 2.
\begin{figure}
\begin{center}
\epsfig{file=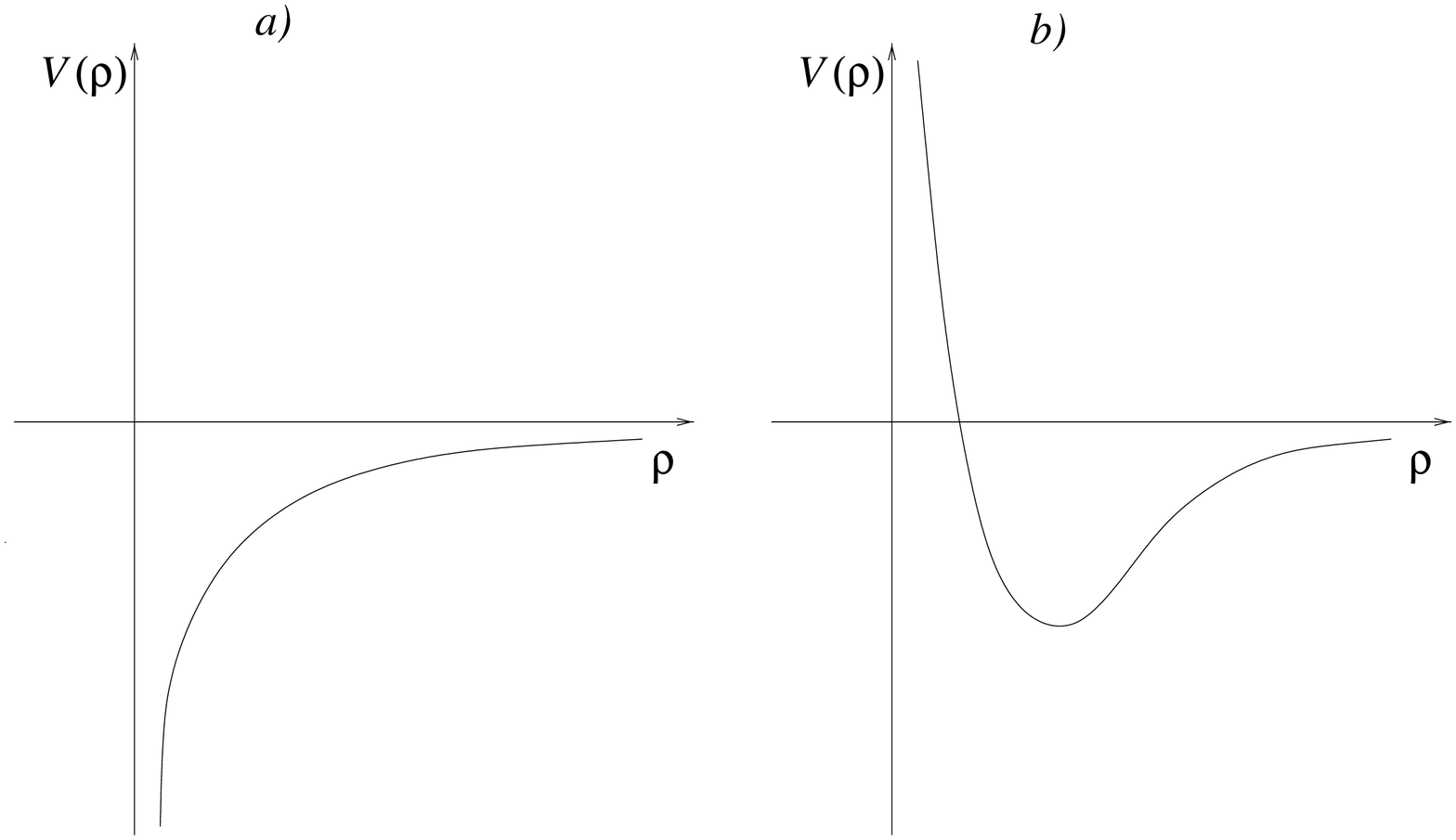,width=12cm,height=6cm}
\end{center}
\caption{ a)  $J<J_{cr}$,      $ $        b)  $J>J_{cr}$ .
\label{J} }
\end{figure}

2. Motion in the Schwarzschild background

To investigate the shell motion in the given Schwarzschild background, we 
start with the same equation as in the preceding subsection, but with zero 
electric charge:  
\begin{eqnarray}
F=-\Delta =1-\frac{2\,G\,m}{r}   , 
\\
\sqrt{\dot {\rho }^{2}+1-\frac{2\,G\,m}{\rho }}-\sqrt{\dot {%
\rho }^{2}+1-\frac{2\,G\,(m+\Delta m)}{\rho }}=\frac{G\,M}{\rho }\sqrt{1+%
\frac{a^{2}}{\rho ^{2}}}  .\nonumber 
\label{Sw} 
\end{eqnarray}
The limit to a given background is the limit at which the total 
energy of the source $E=\Delta m$ is negligible with respect 
to the total energy $m$ of the 
system. Therefore, we should expand Eq.(\ref{Sw}) in $\Delta m$ up to the 
first order; the result is 
\begin{equation}
\Delta m=E=M\sqrt{\stackrel{.}{\rho }^{2}+1-\frac{2\,G\,m}{\rho }}\sqrt{1+%
\frac{a^{2}}{\rho ^{2}}}  . 
\label{SWE}
\end{equation}
The formal difference from Eq.(\ref{atom}) for the relativistic Coulomb 
problem is that now the attractive force is incorporated into the 
kinetic term. 
Equation (\ref{SWE}) can also be considered as describing a shell 
consisting of radially moving particles with variable effective masses 
$M_{eff}=M\sqrt{1+\frac{a^{2}}{\rho ^{2}}}$. 
Introducing the dimensionless quantities 
$\Delta \varepsilon =\frac{\Delta m}{M}$, $\varepsilon =\frac{m}{M}$, 
$x=\frac{\rho }{G\,M}$, $\gamma =\frac{a}{G\,M\,}$\ and the ``potential'' 
$V(x)=\Delta \varepsilon (\dot {x}=0)-1$,\ \ we get 
\begin{equation}
\Delta \varepsilon =\sqrt{1-\frac{2\,\varepsilon }{x}}\sqrt{1+\frac{\gamma 
^{2}}{x^{2}}}=V(x)+1   . 
\label{V(x)}
\end{equation}
Again, we have a critical value of the angular momentum 
$\gamma _{cr}=2\sqrt{3}\varepsilon $\ ( $J_{cr}=2\sqrt{3}G\,m\,\mu $, 
where\ $\mu $\ is the mass of a single particle). The potential for 
$\gamma <\gamma _{cr}$ $(J<J_{cr})$ 
behaves qualitatively like the relativistic Coulomb potential in this case, 
apart from the fact that it now starts from $x=2\,\varepsilon $\ (where it 
takes the value $-1$, the lowest possible value corresponding to 
 $\Delta \varepsilon =0$). 
\begin{figure}
\begin{center}
\epsfig{file=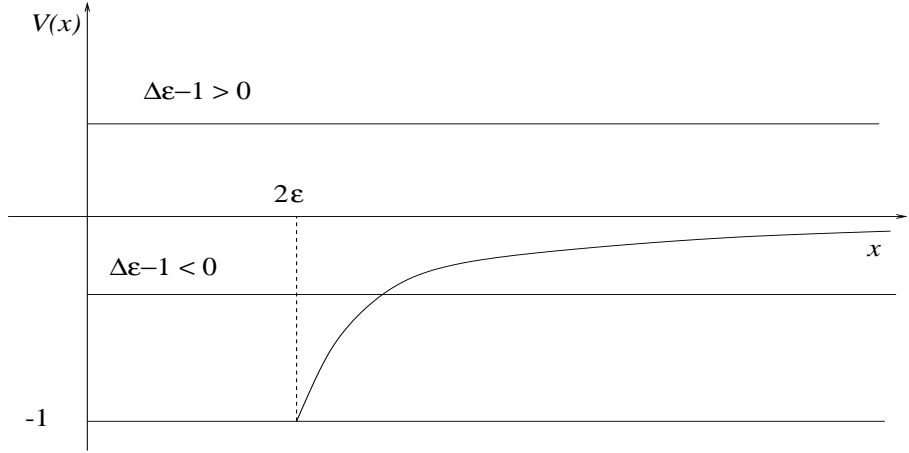,width=12cm,height=6cm}
\end{center}
\caption{$\gamma <\gamma_{cr}=2\sqrt{3} \varepsilon.$ The allowed regions are 
$\Delta \varepsilon \geq V(x)+1$.
\label{V(x)1} }
\end{figure}
Thus, for small enough values of the angular momentum $(\gamma <\gamma _{cr})$ 
we have two possibilities. Either $\Delta \varepsilon >1$, and the shell 
collapses starting from infinity (or, the other way around, it starts from 
the past singularity at $\rho =0$ and escapes to infinity). Or $\Delta 
\varepsilon >1$, and the shell starts from the past singularity at 
$\rho =0$, reaches its maximal radius $\rho _{0}$ at the turning point, 
and then recollapses to the future singularity at $\rho =0$.

When the angular momentum exceeds the critical value, the situation is 
essentially different from the case of the relativistic Coulomb problem. 
Starting from infinity, the potential first goes down to its minimum at $%
x=\frac{6}{\xi }(1-\sqrt{1-\xi })\,\varepsilon $ $\ (\xi =\frac{%
12\,\varepsilon ^{2}}{\gamma ^{2}})$: 
\begin{equation}
V_{\min }=\frac{1}{3}\sqrt{6+\frac{2}{\xi }(1-(1-\xi )^{3\,/\,2})}-1  ,
\label{min}
\end{equation}
grows up to the maximum at $x=\frac{6}{\xi }(1+\sqrt{1-\xi })\,\varepsilon$: 
\begin{equation}
V_{\max }=\frac{1}{3}\sqrt{6+\frac{2}{\xi }(1+(1-\xi )^{3\,/\,2})}-1  ,
\label{max}
\end{equation}
and then again goes down to the minimal possible value $(V=-1)$ at the 
Schwarzschild horizon $(x_{h}=2\,\varepsilon )$. The origin of this new 
feature, the fall off of the potential near the Schwarzschild horizon is that 
in General Relativity any energy gravitates. So, in our case not only the 
bare mass $M$ but also the energy of the angular momentum contribute to the 
gravitational attraction. This is reflected in the fact that the bare 
mass enters the equation only in the combination 
$M\sqrt{1+\frac{a^{2}}{\rho ^{2}}}$. 

Figure 4 shows the behaviour of the potential in the case 
$2\sqrt{3}\varepsilon =\gamma _{cr}<\gamma <\gamma _{1}=4\,\varepsilon $.
\begin{figure}
\begin{center}
\epsfig{file=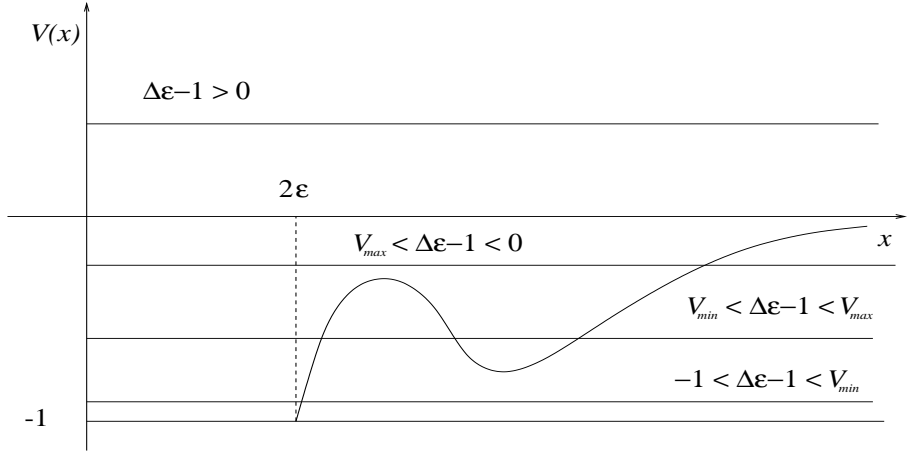,width=12cm,height=6cm}
\end{center}
\caption{$2\sqrt{3}%
\varepsilon =\gamma _{cr}<\gamma <\gamma _{1}=4\,\varepsilon $.
\label{V(x)2} }
\end{figure}
Now we have already four possibilities. Case $1$: $\Delta \varepsilon >1$. 
The unbound motion from infinity to the singularity (and vice versa). 
Case $2 $, $V_{max}+1<\Delta \varepsilon <1$. This is the bound motion, 
the shell starts 
from the past singularity, expands up to the sufficiently large value of 
the radius and recollapses to the future singularity (analogously to the case 
$\Delta \varepsilon <1$\ for $\gamma <\gamma _{cr}$). In this case the shell 
has enough energy $(\Delta \varepsilon )$ to overcome the attraction due to 
a non-zero angular momentum. Case $3$: $V_{\min }+1<\Delta \varepsilon 
<V_{\max }+1$. Here we have two possibilities depending on the initial 
conditions. First, we have orbits, as in the relativistic Coulomb problem. 
Second, we have bound motion with a collapse, but with much smaller value of 
the radius at the turning point (compared with case $2$), and no analogue in 
the Coulomb problem. Case $4$: $0<\Delta \varepsilon <V_{\min }+1$. Only 
bound motion with a collapse. 

For $\gamma =\gamma _{1}=4\,\varepsilon \,$, the value of the potential at 
the maximum is just zero. Figure 5 shows the potential in the case 
$\gamma >\gamma _{1}=4 \varepsilon $. 
\begin{figure}
\begin{center}
\epsfig{file=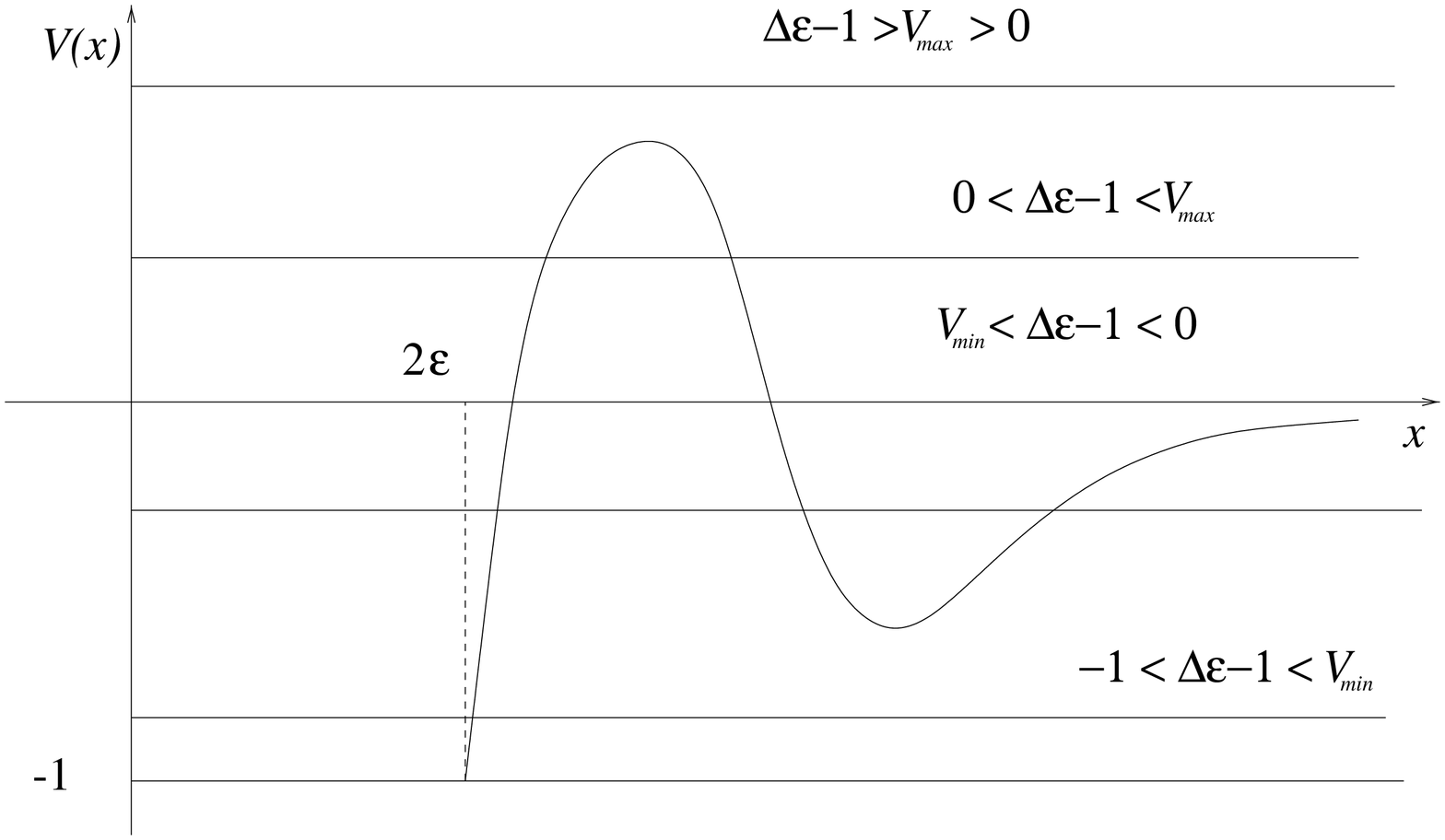,width=12cm,height=6cm}
\end{center}
\caption{ $\gamma >\gamma _{1}=4 \varepsilon $. 
\label{V(x)3} }
\end{figure}
Again, we have four different cases. Case $1$: 
$\Delta \varepsilon >V_{\max }+1>1$. The total energy is large in order 
to overcome the repulsion due to the 
angular momentum, and we have a totally unbound motion $(0<x<\infty )$. Case 
$2$: $1<\Delta \varepsilon <V_{\max }+1$. There are two possibilities, 
depending on the initial conditions. First, as in the Coulomb problem, 
the shell comes 
from infinity, reaches the turning point and goes back to infinity. 
Second, we have bound motion with a collapse, like for $\Delta \varepsilon 
<1 $, (but now $\Delta \varepsilon >1$!). There is no such analogue in 
the Coulomb problem. And, last, Cases $3$ and $4$ for 
$\Delta \varepsilon <1$ are qualitatively the same as Cases $3$ and $4$ 
for lower $(\gamma _{cr}<\gamma <\gamma _{1})$\ angular momentum. 

At the end of this subsection we would like to note that from Eq.(\ref{V(x)})
it is easy to derive the radii for the smallest stable $(\gamma 
=\gamma_{cr})$ and unstable 
$(\gamma \longrightarrow \infty ,\Delta \varepsilon =V_{\max }+1)$ 
circular orbits. They are, of course, the same as for 
geodesics and equal to $\rho _{st}=6 G m$ and  $\rho _{unst}=3 G m$, 
respectively.

3. Self-gravitating shells

Let us now turn to the most important case of self-gravitating shells with 
fill account for a back-reaction of the shell dynamics on the space-time 
structure. We start with the same equation as in the preceding subsection 
(all the notations are also the same) but with $\sigma _{out}=\sigma $ 
(we will not consider here the shells $\sigma _{in}=-1$ because there are no 
solutions in this case with the shell in the $R_{+}$-region which is most 
interesting for us). We get 
\begin{equation}
\sqrt{\dot {x}^{2}+1-\frac{2\,\varepsilon }{x}}-
\sigma \sqrt{\dot{x}^{2}+1-\frac{2\,(\varepsilon +\Delta \varepsilon )}{x}}=
\frac{1}{x}\sqrt{1+\frac{\gamma ^{2}}{x^{2}}} 
\label{m1}
\end{equation}
and 
\begin{equation}
\Delta \varepsilon =\sqrt{\stackrel{.}{x}^{2}+1-
\frac{2\,\varepsilon }{x}}\sqrt{1+\frac{\gamma ^{2}}{x^{2}}}-
\frac{1}{2\,x}\left( 1+\frac{\gamma ^{2}}{x^{2}}\right )  , 
\label{m2}
\end{equation}
while the potential reads now as follows:
\begin{equation}
V(x)=\Delta \varepsilon (\dot {x}=0)-1=\sqrt{1-
\frac{2\,\varepsilon }{x}}\sqrt{1+\frac{\gamma ^{2}}{x^{2}}}-
\frac{1}{2\,x}\left( 1+\frac{\gamma ^{2}}{x^{2}}\right) -1   . 
\label{m3}
\end{equation}
The new feature with respect to the case of a given background is the 
appearance of the self-interaction term in Eq.(\ref{m2}) and, 
as a consequence, the possibility of changing the sign of $\sigma $ in 
Eq.(\ref{m1}). This means that we 
can have not only the shells moving in the $R_{+}$-region (black hole case) 
but also those in the $R_{-}$-region (wormhole case). It is 
clear that in the wormhole case $(\sigma =-1)$ there cannot be unbound 
motion. The change of the sign of $\sigma $ and, hence, the transition from 
the black hole case to the wormhole case takes place when the turning point is 
exactly at the outer horizon, i.e. at 
$x_{0}=2\,(\varepsilon +\Delta \varepsilon )$. It can be shown that, 
for fixed values of both $\varepsilon $
there is exactly one value of $\gamma $ for 
which this condition holds. The other way around, for every 
$\varepsilon $ and $\gamma $ there is only one value of 
$\Delta \varepsilon _{\sigma }$ below which we have a wormhole shell. 
This value of $\Delta \varepsilon _{\sigma }$ becomes negligibly small in 
the limit $\Delta \varepsilon \ll \varepsilon $, i.e. in the 
Schwarzschild background limit, as 
it should be. So, we consider here another limiting case, $\varepsilon =0$, 
when inside the shell there are no other sources of gravitational 
field and the inner space-time is flat (in the general case some expressions 
can also be written but they are completely unreadable). Substituting for $x$ 
the value $x_{\sigma }=2\,(\varepsilon +\Delta \varepsilon )$ in 
Eq.(\ref{m1}) and putting $\dot {x}=0$ we obtain 
\begin{equation}
\sqrt{\frac{\Delta \varepsilon }{\varepsilon +\Delta \varepsilon }}=\frac{1}{%
x_{\sigma }}\sqrt{1+\frac{\gamma ^{2}}{x_{\sigma }^{2}}}  .  
\label{m4} 
\end{equation}
For $\varepsilon =0$ it simplifies to  
\begin{equation}
1=\frac{1}{x_{\sigma }}\sqrt{1+\frac{\gamma ^{2}}{x_{\sigma }^{2}}}  ,  
\label{m5}
\end{equation}
with the following positive solution for $\Delta \varepsilon _{\sigma }$: 
\begin{equation}
\Delta \varepsilon _{\sigma }=\frac{1}{2\sqrt{2}}\sqrt{1+\sqrt{1+4\,\gamma 
^{2}}}  . 
\label{m6} 
\end{equation}
If $\varepsilon \neq 0$, the corresponding value of 
$\Delta \varepsilon _{\sigma }$ is less than for 
$\varepsilon =0$ for given angular momentum $\gamma $. The inequalities 
$\frac{\partial m}{\partial M}>0$ for the black 
hole case and $\frac{\partial m}{\partial M}<0$ for the wormhole case also 
hold. The new feature is the possibility for a wormhole shell to exist even 
for $\Delta \varepsilon >1$, which is completely forbidden for the shells with 
radially moving particles (there $\Delta \varepsilon <\frac{1}{2}$)! 
 
Let us investigate now the behaviour of the potential $V(x)$, Eq.(\ref{m3}). 
The qualitative features are the same for all values of $\varepsilon $, 
so that we will do this only for the simplest (and most tractable) case of 
zero inner mass, that is, we put $\varepsilon =0$. Thus, 
we have for the potential 
\begin{equation}
V=\Delta \varepsilon -1=\sqrt{1+\frac{\gamma ^{2}}{x^{2}}}-
\frac{1}{2\,x}\left( 1+\frac{\gamma ^{2}}{x^{2}}\right ) -1  .
\label{m7}
\end{equation}
Again, as in the case of the Schwarzschild background, there exists some 
critical value of the angular momentum (in our dimensionless notation it is 
$\gamma _{cr}$) below which the potential looks as follows. 
\begin{figure}
\begin{center}
\epsfig{file=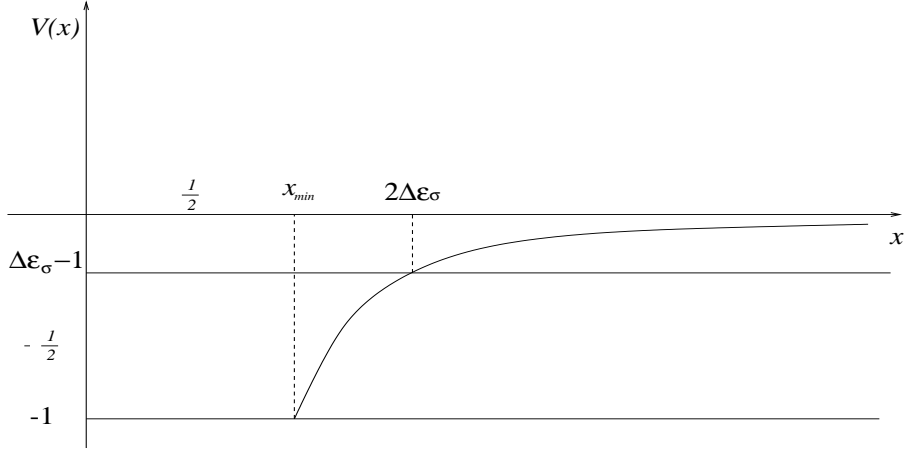,width=12cm,height=6cm}
\end{center}
\caption{$\gamma < \gamma_{cr}$.
\label{V(x)4} }
\end{figure}
It starts from the minimal possible value $V=-1$ (corresponding to 
$\Delta \varepsilon =0$) at 
$x_{\min }=\frac{1}{2\sqrt{2}}\sqrt{1+\sqrt{1+16\gamma ^{2}}}$, 
crosses the horizontal line $(\Delta \varepsilon _{\sigma }-1)$ 
at $x_{\sigma }=2\Delta \varepsilon _{\sigma }=\frac{1}{\sqrt{2}}\sqrt{1+
\sqrt{1+4\gamma ^{2}}}$, and then increases to zero at infinity. 
Below the line $(\Delta \varepsilon _{\sigma }-1)$ we have wormhole 
solutions, and above it 
there are only collapsing shells that start either from some turning point 
(for $\Delta \varepsilon _{\sigma }<\Delta \varepsilon <1$) or from infinity 
(for $\Delta \varepsilon >1$) at the $R_{+}$-region. 

To calculate the value of the critical angular momentum, we must solve (as 
in the Schwarzschild case) the following set of equations 

\begin{eqnarray} 
\Delta \varepsilon  &=&\sqrt{1+\frac{\gamma ^{2}}{x^{2}}}-
\frac{1}{2\,x}\left( 1+\frac{\gamma ^{2}}{x^{2}}\right) =V(x)+1 
\nonumber\\
V^{ \prime}(x) &=&0=-\frac{\gamma ^{2}}{x^{3}\sqrt{1+
\frac{\gamma ^{2}}{x^{2}}}}+\frac{1}{2\,x^{2}}\left( 1+
\frac{3\,\gamma ^{2}}{x^{2}}\right) \\
V^{ \prime \prime}(x)&=&0=-\frac{\gamma ^{4}}{x^{6}(1+
\frac{\gamma ^{2}}{x^{2}})
\sqrt{1+\frac{\gamma ^{2}}{x^{2}}}}-
\frac{3\,\gamma ^{2}}{x^{4}\sqrt{1+
\frac{\gamma ^{2}}{x^{2}}}}-
\frac{1}{x^{3}}\left( 1 + \frac{6\gamma ^{2}}{x^{2}}\right)   .\nonumber
\label{m8}
\end{eqnarray}
Replacing the combination $(\frac{\gamma ^{2}}{x^{2}})$ by 
$\xi $\ $(>0)$\ everywhere in the two last equations, we obtain a quadratic 
equation for $\xi $\ with the positive solutions  
\begin{equation}
\xi =\frac{-3+\sqrt{33}}{12}  .
\label{m9}
\end{equation}
By a straightforward calculation we find 
\begin{equation}
x_{cr}=\frac{(9+\sqrt{33})^{3/2}}{8\sqrt{3}}\approx 4.0859 
\label{m10}
\end{equation}
\begin{equation}
\gamma _{cr}^{2}=\frac{(1+\sqrt{33})(9+\sqrt{33})^{2}}{3\cdot 2^{7}}\approx
3.8184 
\label{m11} 
\end{equation}
\begin{equation}
\Delta \varepsilon _{cr}=\frac{(5+\sqrt{33})\sqrt{9-\sqrt{33}}}{24}\approx 
0.8077 .
\label{m12} 
\end{equation}
For $\gamma >\gamma _{cr}$ the potential is qualitatively the same as in the 
Schwarzschild case, but now we may have coexisting wormhole shells and 
``ordinary'' shells with the same parameters (but different initial 
conditions), the latter having either elliptic trajectories, as shown in 
Fig. 7 or hyperbolic ones (Fig. 8) (note the line $(\Delta \varepsilon 
_{\sigma }-1)$\ never reaches the maximum of the potential). 
\begin{figure}
\begin{center}
\epsfig{file=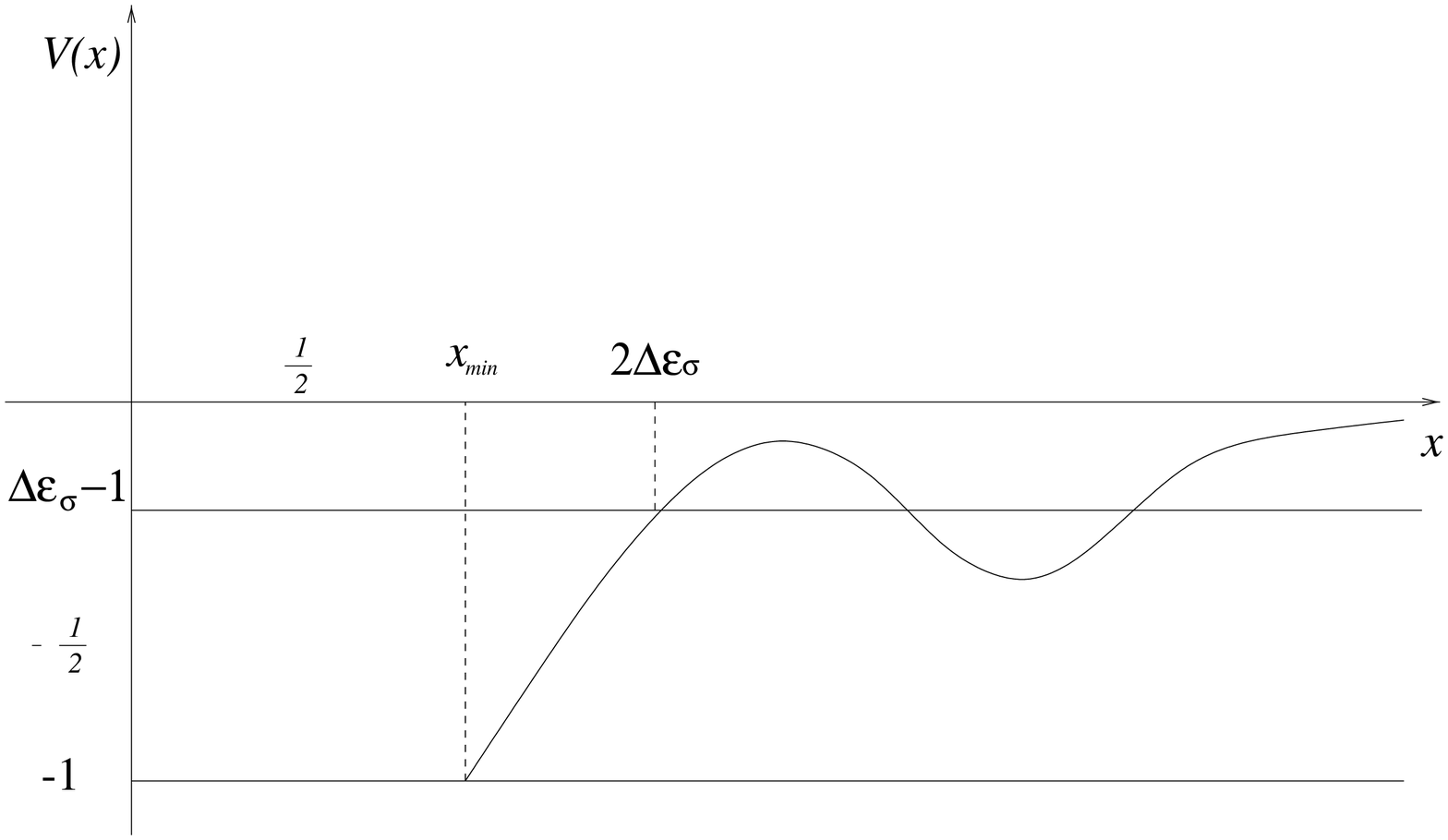,width=12cm,height=6cm}
\end{center}
\caption{$\gamma > \gamma_{cr}$.
\label{V(x)5} }
\end{figure}

\begin{figure}
\begin{center}
\epsfig{file=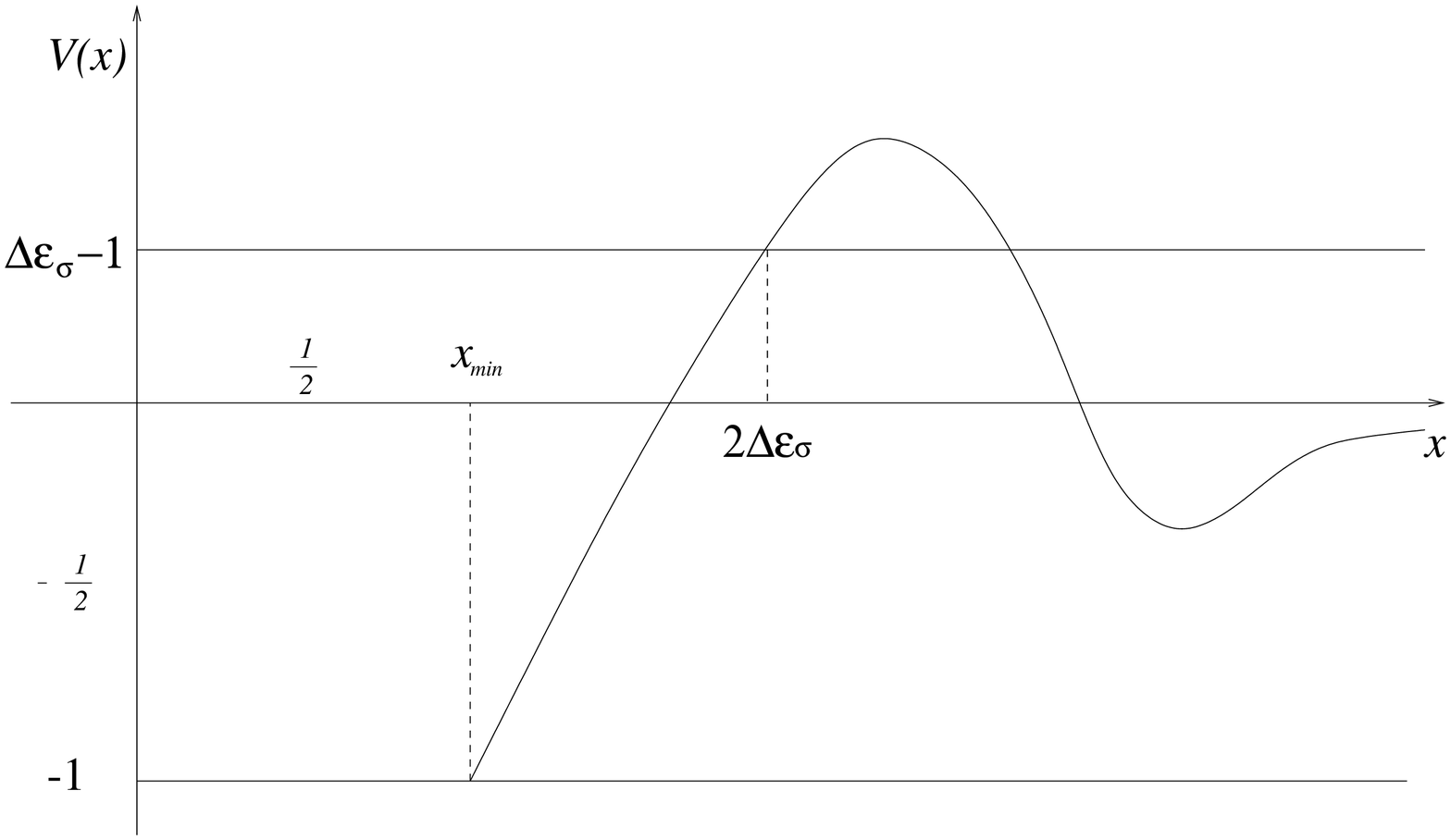,width=12cm,height=6cm}
\end{center}
\caption{$\gamma > \gamma_{cr}$.
\label{V(x)6} }
\end{figure}

\section{Shells with massless particles}

In this section we study the shells with orbiting massless particles. It 
should be stressed that althought particles are moving along null curves the 
shell itself is not null since a projection of a particle's velocity onto a 
radial coordinate line results in a time-like curve. The corresponding limit 
is therefore very simple and straightforward. Indeed, let us look 
at our original equation (we confine ourselves to the $\sigma _{in}=1$ case):  
\begin{equation}
\sqrt{\dot {\rho }^{2}+1-\frac{2\,G\,m}{\rho }}-\sigma \sqrt{%
\dot {\rho }^{2}+1-\frac{2\,G\,(m+\Delta m)}{\rho }}=\frac{G\,M}{\rho 
}\sqrt{1+\frac{a^{2}}{\rho ^{2}}} 
\label{n1}
\end{equation}
and remember that the specific angular momentum $a=\frac{J}{\mu }$ and $%
M=N\,\mu $, where $J$ is an angular momentum of a single particle, $\mu $ is 
its mass and $N$ is a number of particles in the shell. Now, taking the 
limit $\mu \longrightarrow 0$\ we get 
\begin{equation}
\sqrt{\dot {\rho }^{2}+1-\frac{2\,G\,m}{\rho }}-\sigma \sqrt{%
\dot {\rho }^{2}+1-\frac{2\,G\,(m+\Delta m)}{\rho }}=\frac{G\,L}{\rho 
^{2}}  . 
\label{n2}
\end{equation}
Here we introduced a ``total'' angular momentum $L=N\,\mu $, although the real 
total angular momentum of our shell is, by construction, zero. Making the 
radius and masses dimensionless, $\rho =l_{pl}\,y=\sqrt{G}y$, $m=m_{pl}\nu =%
\frac{\nu }{\sqrt{G}}$ ($L$ is already dimensionless) we have 
\begin{equation}
\sqrt{\dot {y}^{2}+1-\frac{2\,\nu }{y}}-\sigma \sqrt{\dot {y}%
^{2}+1-\frac{2\,(\nu +\Delta \nu )}{y}}=\frac{L}{y^{2}}  . 
\label{n3} 
\end{equation}
Note that for $y\longrightarrow \infty$ the rapidity $\left| \dot {y}\right|$ 
(a proper time velocity) tends to infinity as 
$\left| \dot {y}\right| \sim \sqrt{\frac{\Delta \nu }{L}y}$, that is, 
the shell is accelerated to the speed of light. 
 
For the sake of simplicity, here we, again, consider only two limiting 
cases, namely, the limit of the Schwarzschild background 
$\Delta \nu \ll \nu$, and the case $\nu =0$ (no other sources inside 
the shell).

In the first case $(\Delta \nu \ll \nu )$ we have 
\begin{equation}
\Delta \nu =\frac{L}{y}\sqrt{\dot {y}^{2}+1-\frac{2\,\nu }{y}}\, , 
\label{n4}
\end{equation}
and, introducing a potential $U(y)=\Delta \nu (\dot {y}=0)$, 
\begin{equation}
U(y)=\frac{L}{y}\sqrt{1-\frac{2\,\nu }{y}}  . 
\label{n5}
\end{equation}
This potential has a maximum at $y=3\,\nu $ with 
$U_{\max }=\frac{L}{3\sqrt{3}\nu }$, as is shown in Fig.9. 
\begin{figure}
\begin{center}
\epsfig{file=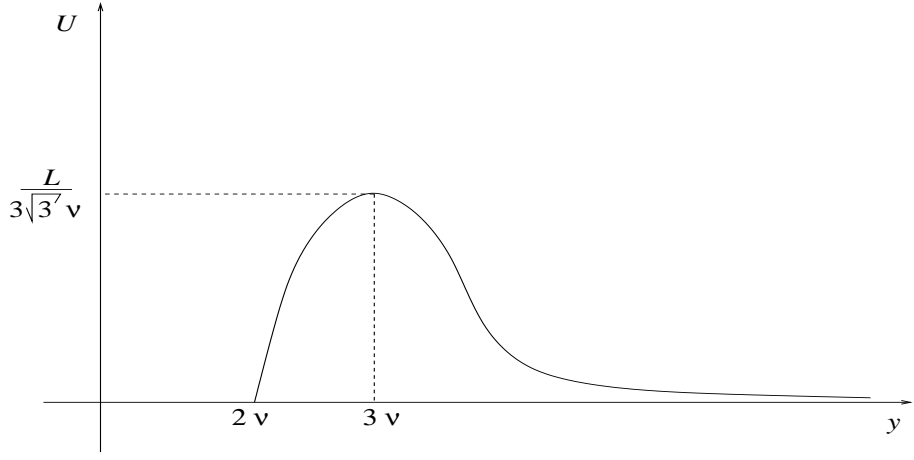,width=12cm,height=6cm}
\end{center}
\caption{$\Delta \nu \ll \nu$.
\label{U1} }
\end{figure}
We see that the higher the angular momentum $L$, the higher the maximal value 
$U_{\max }$, and the larger the black hole mass $\nu $, the lower the maximum. 

In the second case of inner mass $\nu =0$ we have for the turning 
points 
\begin{equation}
1-\sigma \sqrt{1-\frac{2\,\Delta \nu }{y}}=\frac{L}{y^{2}}  , 
\label{n6}
\end{equation}
\begin{equation}
\Delta \nu =\frac{L}{y}\left( 1-\frac{L}{2\,y^{2}}\right) =U(y) . 
\label{n7}
\end{equation}
The changing in the sign of $\sigma $ occurs for $y=2\,\Delta \nu $: 
\begin{eqnarray}
\Delta \nu _{\sigma }=\frac{1}{2}\sqrt{L}  , \nonumber\\ 
y_{\sigma }=\sqrt{L}  , 
\label{n8}
\end{eqnarray}
while the maximum of the potential is achieved at $y=\sqrt{\frac{3}{2}L}$, 
\begin{equation}
U_{\max }=\left( \frac{2}{3}\right) ^{3/2}\sqrt{L}>\Delta \nu _{\sigma }  . 
\label{n9}
\end{equation}
The potential is shown in Fig. 10.
\begin{figure}
\begin{center}
\epsfig{file=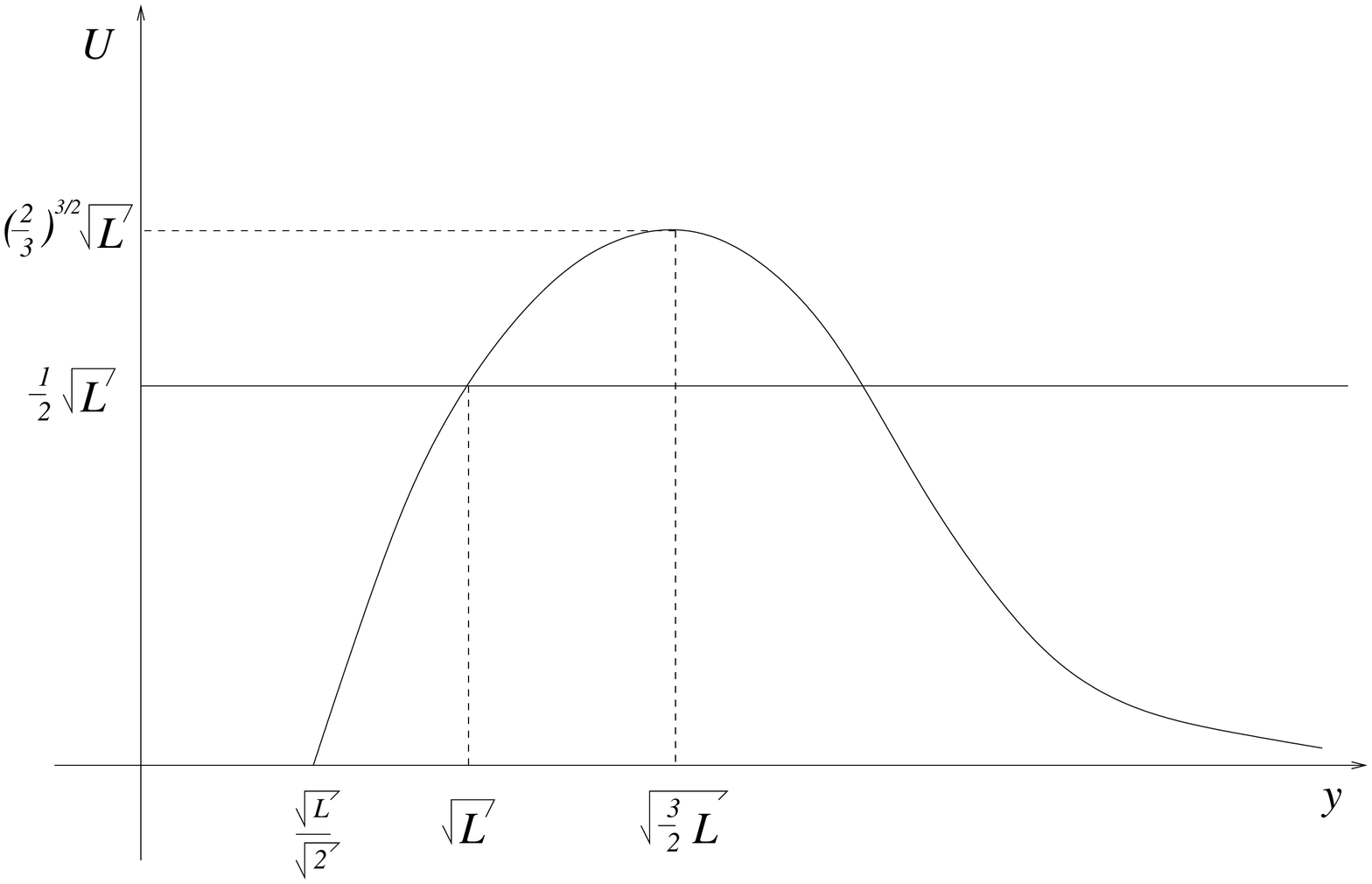,width=12cm,height=6cm}
\end{center}
\caption{$\nu = 0$.
\label{U2} }
\end{figure}
Thus, above the maximum, the shells coming from infinity collapse; the 
less energetic shells have a reflexion point, i.e. they start to infinity, 
shrink up to the minimal radius and then accelerate back to infinity. At 
the same time there coexist shells with the same energy (but with different 
initial conditions) that start from the past singularities at $\rho =0$, 
expand to the maximal radius, and then recollapse to the future singularity 
at $\rho =0$. If their energy is above $\Delta \nu _{\sigma }$, they live in 
the $R_{+}$-region (the black hole case), while if the energy is lower they 
form wormholes. In the extreme limit $\Delta \nu =0$, such a wormhole closes 
up, and we have a closed Universe with only one thin shell with the 
constituent particles moving with the speed of light.

\section{Discussion}

In this paper we investigated the self-gravitating, spherically symmetric 
thin shells constructed of orbiting particles. Is there anything new 
with respect to 
the well-known and well-understood geodesic motion in the Schwarzschild 
background? 

Yes, the self-consistent solution of Einstein's equations for a gravitational 
field, together with the equations for matter fields (dust particles in our 
case), leads to two new fatures. 

First, we found that there exists a minimal possible critical value for an 
angular momentum at which it becomes possible for particles to have an 
elliptic orbit. Such a minimum occurs for the shell inside which there is no 
other sources of gravitational field, and the inner space-time is, therefore, 
flat. In our dimensionless notation it is expressed as  
$\gamma _{cr}\simeq 1.95$. In the conventional units it becomes 
\begin{equation}
J_{cr}=N\,\left( \frac{\mu }{m_{pl}}\right) ^{2}\gamma _{cr}  ,
\label{d1}
\end{equation}
where $m_{Pl}\approx 10^{-5}$ gr is the Planckian mass, $\mu $ is the mass of 
an individual particle, and $N$ is the number of particles in the shell. Of 
course, this value is negligibly small for any atomic or subatomic system. 
But for the other extreme situation, say, for stellar nebula, it may be 
rather important. Indeed, the stars in this case moving along elliptic 
orbits around a common centre of mass can be approximated by a set of ``thin'' 
spherically symmetric shells with orbiting dust particles. For the 
innermost layer we have just the minimal critical angular momentum. The 
ratio $(\frac{\mu }{m_{Pl}})$ is now huge, as the number of stars 
could be. But how about the value $J_{cr}$? Is it large or small? To 
understand this let us calculate the angular velocity $\omega $ of the 
particles moving along circular orbit with $J=J_{cr}$ and radius 
$\rho =\rho _{cr}$. Using the expression for the four-velocities at the very 
end of Section III, we easily obtain 
\begin{eqnarray}
\omega _{cr}=\frac{a_{cr}}{\rho _{cr}^{2}\sqrt{1+
\frac{a^{2}}{\rho ^{2}}}}  , \nonumber\\ 
a_{cr}=\frac{J_{cr}}{\mu }  .
\label{d2}
\end{eqnarray}
Thus, the linear velocity $V$ does not depend on both mass $\mu $ and number  
$N$ of particles and reads 
\begin{equation}
V_{cr}=\frac{a_{cr}}{\rho _{cr}\sqrt{1+\frac{a^{2}}{\rho ^{2}}}}=\frac{%
\gamma _{cr}}{x_{cr}\sqrt{1+\frac{\gamma _{cr}^{2}}{x_{cr}^{2}}}}=\left( 
\frac{\sqrt{33}-3}{\sqrt{33}+9}\right) ^{\frac{1}{2}}\approx 0.4  . 
\label{d3}
\end{equation}
It can be shown that the linear velocity of the circular orbits (at the 
minimum of the potential) goes down with increasing $\gamma $. So, the 
maximal possible linear velocity for the stable circular orbit is 40\%\ of 
the speed of light (the latter is just $1$ in the chosen units). This limit 
should be compared with the corresponding limit for a geodesic motion in the 
Schwarzschild background, where $\frac{\gamma _{cr}}{x_{cr}}=\frac{1}{\sqrt{3%
}}$ and $v_{cr}=\frac{1}{2}$. Thus, the account for back-reaction of the 
matter fields on the gravitational field leads in our case to a more severe 
restriction of the maximal possible velocity for the circular orbits. Maybe 
this will help us to understand how black holes could be formed in 
the galactic nuclei.

The second new features is the wormhole solutions. The necessity to take 
into account these (rather unusual) solutions becomes crucial in quantum 
theory. It was shown in \cite{bbn} that the very existence of the wormhole 
$R_{-}$
-region requires a new quantum number for ``good'' wave functions. For 
example, for simplest bound motions, we need two quantum numbers (with only 
one in the conventional quantum mechanics), and, what is more surprising, we 
have a quantum number for an unbound motion, in particular for radiating 
light-like (null) shells. It seems that it is a radiation spectrum that is 
responsible for a quantum black hole mass spectrum; and  structure of  
matter field energy levels is responsible for the black hole entropy. 
To get a highly degenerate black hole mass spectrum (and, thus, a huge 
entropy) we need quite rich a structure of both radiation and matter source 
spectra.
The features of the potential considered in this paper seem 
sufficiently rich for such purposes. We mean that in the case of the 
orbiting particles we have not only the existence of the wormhole regions 
but also the coexistence of both the elliptic or hyperbolic orbits and 
wormhole solutions for the same values of the parameters (but different 
initial conditions). Such a coexistence will lead to a double-splitting of 
the energy levels as in the famous quantum mechanical double-well problem.

\section*{Acknowledgements}

First of all, we would like to thank G. Bisnovaty-Kogan. The authors 
appreciate the financial support of the Russian Foundation for Basic 
Research (grant 99-02-18524-a). Victor Berezin is greatly indebted to 
Michelle Mazerand, Nanie Perrin and Suzy Vascotto for their help in 
preparing the manuscript and warm hospitality during his stay in the 
TH-Division of CERN.

\end{document}